\documentclass[manuscript]{aastex61}

\newcommand\aastex{AAS\TeX}

\submitjournal{PASP}

\shorttitle{\aastex\  Arp 2 Variable Stars}
\shortauthors{Pritzl et al.}

\begin{document}

\title{Variable Stars in Sagittarius Globular Clusters I. Arp~2}

\correspondingauthor{Barton J. Pritzl}
\email{pritzlb@uwosh.edu}

\author{Barton J. Pritzl}
\affiliation{Department of Physics \& Astronomy, University of Wisconsin Oshkosh, Oshkosh, WI 54901 USA}

\author{Thomas C. Gehrman}
\affiliation{Department of Physics \& Astronomy, University of Wisconsin Oshkosh, Oshkosh, WI 54901 USA}
\affiliation{Minnesota State University Mankato, Mankato, MN 56001}

\author{Ricardo Salinas}
\affiliation{Gemini Observatory, Casilla 603, La Serena, Chile}

\author{M\'{a}rcio Catelan}
\affiliation{Instituto de Astrof\'{i}sica, Pontificia Universidad Cat\'{o}lica de Chile, Av.\ Vicu\~{n}a Mackenna 4860, 782-0436 Macul, Santiago Chile}
\affiliation{Millennium Institute of Astrophysics, Santiago, Chile}

\author{Horace A. Smith}
\affiliation{Department of Physics \& Astronomy, Michigan State University, East Lansing, MI 48824, USA}

\author{Jura Borissova}
\affiliation{Instituto de F\'{i}sica y Astronom\'{i}a, Universidad de Valpara\'{i}so, Av. Gran Breta\~{n}a 1111, Playa Ancha, Casilla 5030, Chile}
\affiliation{Instituto de Astrof\'{i}sica, Pontificia Universidad Cat\'{o}lica de Chile, Av.\ Vicu\~{n}a Mackenna 4860, 782-0436 Macul, Santiago Chile}

\begin{abstract}

We present new photometry and analysis of the twelve variable stars (nine RR~Lyrae, three SX~Phoenicis) belonging to the Sagittarius globular cluster Arp~2. Of the nine RR~Lyrae stars in the cluster, eight are RRab and one is RRc. From the RRab stars, we determined a mean period of $\langle P_{ab}\rangle=0.581\pm0.047$~days, where the error is the standard error of the mean. This places Arp~2 at the border between the Oosterhoff~I and Oosterhoff-Int clusters. Using the $V$-band data from the RR~Lyrae stars, a distance modulus of $(m-M)_0=17.24\pm0.17$ was determined. From the $I$-band data, we found $(m-M)_0=17.34\pm0.07$. We also used the SX~Phoenicis variables to determine a distance modulus of $(m-M)_0=17.27\pm0.04$. Color excesses were determined from the RR Lyrae light curves using both the ($B-V$) and ($V-I$) colors. The mean reddening values were in line with or were a little higher than those found in the literature. Both methods indicated star-to-star variability in the reddening toward Arp~2. Of the nine RR~Lyrae stars, seven were flagged as variables by Gaia, with three having periods determined. We used the Gaia data to investigate the membership of the seven Gaia RR~Lyrae. Although Arp~2 is too distant for reliable Gaia parallax, the current data do not exclude any of the variables discussed in this paper from being members of Arp~2.

\end{abstract}


\section{Introduction} \label{sec:intro}

The discovery of the Sagittarius (Sgr) dwarf spheroidal galaxy (Ibata et al.\ 1994) has allowed astronomers to study an extragalactic system up-close. Soon after, it was observed that the Sgr dwarf galaxy had four globular clusters (GCs) associated with it, M54, Arp~2, Terzan~7, and Terzan~8 (Ibata et al.\ 1995). Since then, a number of additional GCs have been suggested as members or former members. It has been thought that large galaxies like the Milky Way were formed in part through mergers with smaller galaxies (e.g., Font et al.\ 2011). Studying a system like the Sgr dwarf galaxy allows us to better understand this process.

The Milky Way GC system has been shown to have a dichotomy of properties when the properties of the RR~Lyrae (RRL) stars within them are compared. Oosterhoff (1939) found that metal-poor GCs tend to have RRL stars with a longer mean period compared to metal-rich GCs. There is a gap between the two groups (metal-rich -- Oosterhoff I and metal-poor -- Oosterhoff II) where Milky Way GCs are able to produce few if any RRL stars. It has been found that GCs that belong to extragalactic systems can produce RRL within the Oosterhoff gap (see Catelan \& Smith 2015 for summary). This is why studying systems like the Sgr dwarf GCs is very important.

This paper is a follow-up to Salinas et al.\ (2005), which did an initial search for variable stars in Arp~2, Terzan~8, NGC~5634, and Palomar~12. That paper used an image subtraction technique to detect the variables and determine their periods, but did not provide magnitudes and other information about the variable stars. Here we begin a series of papers taking a detailed look at the variable star populations within each cluster as well as what they tell us about the nature of the parent cluster and the Sgr dwarf galaxy.

Arp~2, Terzan~7, Terzan~8, and Palomar~12 have all been confirmed to be associated with the Sgr dwarf galaxy through a proper motion study (Sohn et al.\ 2018). Arp~2 is an interesting cluster because it appears to be relatively young while still metal-poor. It has been shown to be about 3-4 Gyr younger than older GCs, but about 1-2 Gyr older than the youngest GCs (Layden \& Sarajedini 2000; Carraro, Zinn, \& Moni Bidin 2007). Mottini, Wallerstein, \& McWilliam (2008) found the metallicity to be [FeI/H]$=-1.77\pm0.04$ with [Fe/H]$=-1.83\pm0.07$ when the FeII lines were included. Also, they found [$\alpha$/Fe]$=0.31\pm0.11$, which is not very different from Galactic GCs. Salinas et al.\ (2005) found a total of nine RRL and three SX~Phoenicis (SX~Phe) stars within Arp~2. In the following we detail the observations and reductions, the properties of the variable stars, and how Arp~2 compares to the other Sgr GCs and Milky Way GCs.

\section{Observations and Reductions} \label{sec:observations}

Arp~2 was observed, along with the Terzan~8, NGC~5634, and Palomar~12, from June 28 - July 4, 2003 using the Danish Faint Object Spectrograph and Camera on the Danish 1.54-m telescope at La Silla, Chile. The field-of-view is 13.7\arcmin x 13.7\arcmin. In total 42 images were taken in the Johnson $B$ filter (exposure times between 250 and 650 seconds), 45 in the Johnson $V$ filter (exposure times between 100 and 350 seconds), and 21 in the Cousins $I$ filter (exposure times between 120 and 200 seconds). The seeing ranged from 0.7\arcsec to 3.3\arcsec with an average seeing of 1.3\arcsec. 

Images were bias-subtracted and flat-field corrected using the standard routines in IRAF\footnote{IRAF is distributed by the National Optical Astronomy Observatory, which is operated by the Association of Universities for Research in Astronomy, Inc., under cooperative agreement with the National Science Foundation.}. Point-spread functions were created for each image using the DAOPHOT~II stand-alone program (Stetson 1987). Magnitudes were obtained using the ALLFRAME routine (Stetson 1994). The magnitudes were calibrated using 75 local standard stars from the Stetson online database (Stetson 2000). When compared to another study of Arp~2 we find our magnitudes are brighter by 0.04 in $B$ and $V$ on average using 180 stars and 200 stars, respectively, from Buonanno et al. (1995). The Buonanno et al.\ data were calibrated using ten stars within M67 and five stars in a Landolt (1992) field. When compared to the study by Carraro \& Seleznev (2011), we find a greater difference. Using 30 stars in $B$ and $V$ from their lists of horizontal branch and blue stragglers, we find our magnitudes are fainter on average by 0.21 in $B$ and 0.23 in $V$. The Carraro \& Seleznev data were calibrated using two Landolt fields. Comparing these two studies, we would see the Buonanno et al.\ data would also be much fainter than the Carraro \& Seleznev study. Anderson et al.\ (2008) also looked at Arp~2 as part of a survey of globular clusters. While we were not able to match our coordinate system with their coordinates, we plotted their $V$,$V-I$ color-magnitude diagram (CMD) against ours. The CMD trend lines appear to be in agreement to within a few hundredths of a magnitude.

The variability index from ALLFRAME was used to identify candidate variable stars. Each candidate was examined by eye using the phase dispersion minimization technique (PDM, Stellingwerf 1978) in IRAF. Periods were determined for each variable star and for each filter according to which one gave the least amount of scatter to the light curve. Because of the larger number of data points, the final periods were determined using the $B$ and $V$ data. In the majority of cases, the periods in both filters were consistent with each other and well within their uncertainties. When a slight difference was found, the periods from the two filters were averaged.

\section{RR Lyrae Stars} \label{sec:rrl}

Valenti (2001) found four RRL in Arp~2. Salinas et al.\ (2005) found those four RRL and an additional five RRL using an image-subtraction technique. In our photometry we were able to rediscover the nine RRL stars. Figure~1 shows the location of the RRL stars, along with the SX~Phe stars, within the CMDs. There is significant field contamination within the CMDs. We used a tidal radius of 4.61\arcmin$\pm$0.30\arcmin from Salinas et al.\ (2012) to distinguish between those stars that likely belong to the cluster and those that likely belong only to the field. There are two variables, V5 and V7, that fall outside of the tidal radius. However, a comparison of the inner and outer CMDs shows that there do appear to be some horizontal branch and blue straggler stars in the outer CMD. The variable V7 is seen to have colors that are shifted toward the red compared to the other RRL stars. Visual inspection does not appear to show any blending with another star. Some of the issues may stem from poor light curve fits to the data given the maximum was missed in each filter. More data would be needed to accurately determine if this is an issue with the photometry or if V7 may not be a member of Arp~2. In the following we will assume that V5 and V7 are a part of the Arp~2 cluster, except where indicated otherwise. The issue of cluster membership will be addressed in Section~3.1.

\begin{figure}
     \includegraphics[width=\columnwidth]{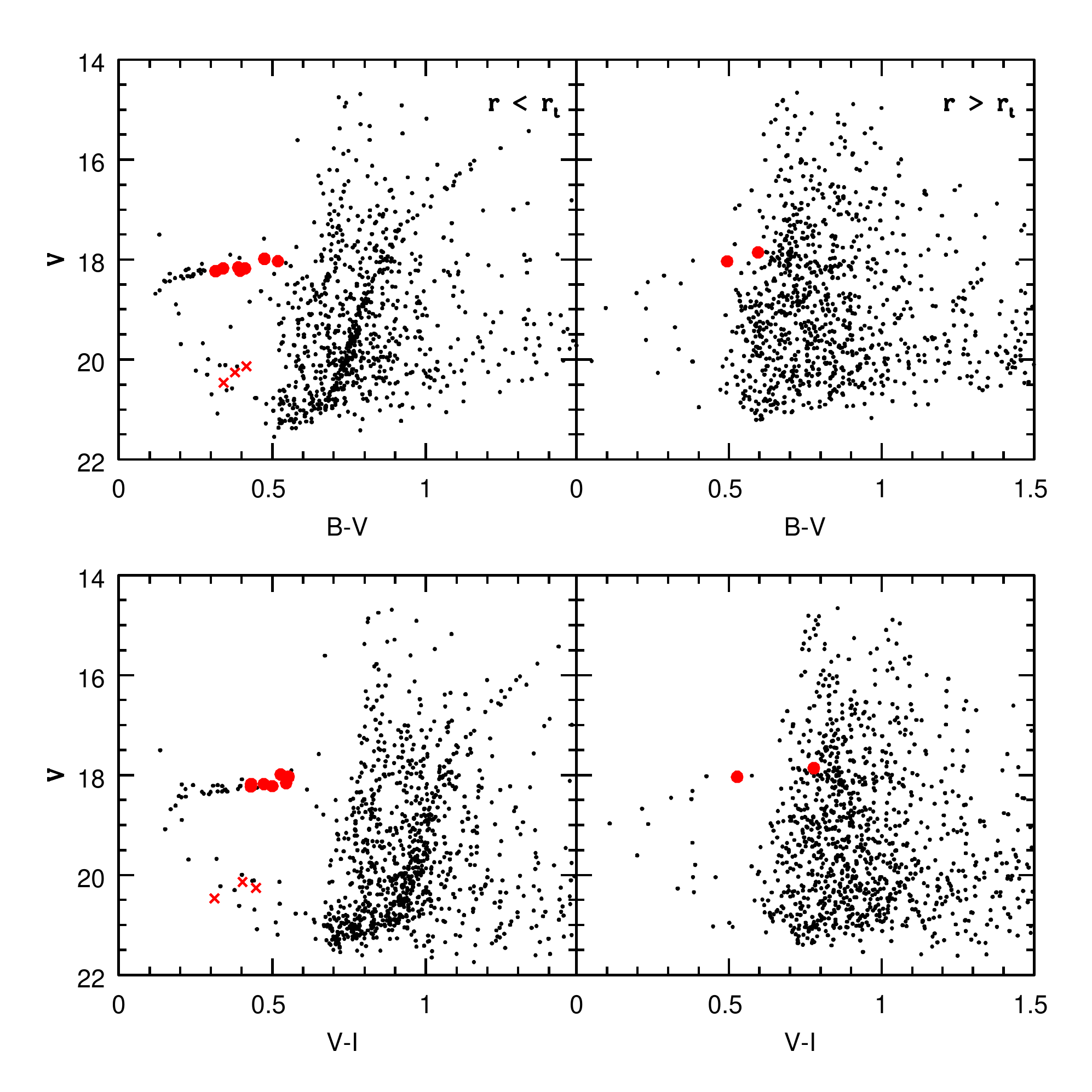}
     \caption{Color-magnitude diagrams for Arp~2 showing $V$ versus $B-V$ in the upper two panels and versus $V-I$ in the lower two panels. As indicated, the panels on the left are those stars found within the Arp~2 tidal radius while the panels on the right are those outside of the tidal radius (4.61\arcmin$\pm$0.30\arcmin, Salinas et al.\ 2012). RR~Lyrae stars are shown as large red points, whereas the SX~Phe stars are shown as crosses.}
\end{figure}

Table~1 lists the properties of the RRL stars. The intensity-weighted magnitudes were determined through a routine that uses model RRL light curves to fit the data (A. Layden, private communications). A sample of the light curves is given in Figure~2. All of the light curves can been seen in the electronic edition of the paper. As noted above, even though a routine that produced a fit to the entire light curve was used, there are a couple of cases where the peak of the light curve was not observed due to lack of phase coverage. This may have affected the calculated magnitude, as is the case with V7. We do not include poor fits to the data when plotting the light curves as can be seen in Figure~2. We also attempted to use Fourier analysis to examine the light curves using the method outlined in Kov\'{a}cs \& Kupi (2007). For a couple of the stars, we were able to determine [Fe/H] values, but the gaps in the light curves resulted in inadequate fits to determine properties for most of the RRL stars. Given the small number of adequate fits, we do not include the Fourier analysis in this paper. Figure~3 shows the location of the variable stars in Arp~2. In comparing the periods we derived to those determined by Salinas et al.\ (2005) the only notable change was for V2 with a period change of 0.01. Most of the other periods differed within the uncertainties. It should be noted that V2 stands out from among the other RRL stars. It has a notably long period and it is brighter than the other RRL. Its location in the field does not give any indication that it would be outside of Arp~2. In the next subsection we discuss the possible membership of the RRL stars.

 \begin{table*}
  \caption{Properties of the Arp~2 Variable Stars}             
  \label{table:1}      
  \centering          
  \begin{tabular}{c c c c c c c c c c l}     
  \hline\hline       
  ID  & R.A.      & Dec.       & Period & $\langle B\rangle$ & $\langle V\rangle$ & $\langle I\rangle$ & $A_{B}$ & $A_{V}$ & $A_{I}$ & Comments\\
       & J2000.0 & J2000.0 & (days) & \\ 
  \hline                    
     1 & 19:28:36.4 & $-30$:20:56.2 & 0.5684$\pm$0.0007     & 18.59 & 18.18 & 17.71 & 1.49 & 1.15 & 0.42 & RRab \\  
     2 & 19:28:39.4 & $-30$:20:07.0 & 0.811$\pm$0.001         & 18.55 & 18.03 & 17.48 & 0.64 & 0.49 & 0.33 & RRab \\
     3 & 19:28:56.9 & $-30$:21:29.6 & 0.565$\pm$0.001         & 18.46 & 17.99 & 17.46 & 1.14 & 0.90 & 0.29 & RRab \\
     4 & 19:29:01.5 & $-30$:20:56.1 & 0.458$\pm$0.001         & 18.55 & 18.23 & 17.80 & 1.41 & 1.06 & 0.30 & RRab \\
     5 & 19:28:56.3 & $-30$:26:32.0 & 0.760$\pm$0.001         & 18.53 & 18.04 & 17.51 & 0.70 & 0.59 & 0.34 & RRab \\
     6 & 19:28:30.2 & $-30$:23:11.3 & 0.4472$\pm$0.0005     & 18.61 & 18.22 & 17.72 & 1.16 & 0.90 & 0.64 & RRab \\
     7 & 19:29:04.8 & $-30$:16:04.6 & 0.527$\pm$0.002         & 18.45 & 17.86 & 17.08 & 0.66 & 0.56 & 0.07 & RRab \\
     8 & 19:28:44.9 & $-30$:21:54.7 & 0.2921$\pm$0.0007     & 18.52 & 18.18 & 17.75 & 0.62 & 0.49 & 0.27 & RRc \\
     9 & 19:28:52.3 & $-30$:22.10.7 & 0.515$\pm$0.001         & 18.55 & 18.16 & 17.62 & 1.60 & 1.28 & 0.78 & RRab \\
   10 & 19:28:52.1 & $-30$:23.13.3 & 0.04738$\pm$0.00003 & 20.81 & 20.47 & 20.16 & 0.55 & 0.50 & 0.19 & SX~Phe \\
   11 & 19:28:45.6 & $-30$:20.05.8 & 0.06121$\pm$0.00003 & 20.56 & 20.14 & 19.74 & 0.36 & 0.35 & 0.22 & SX~Phe \\
   12 & 19:28:42.4 & $-30$:25.15.9 & 0.06040$\pm$0.00003 & 20.64 & 20.26 & 19.81 & 0.36 & 0.28 & 0.25 & SX~Phe \\
  \hline                  
  \end{tabular}
  \end{table*}

\begin{figure}
     \includegraphics[width=\columnwidth]{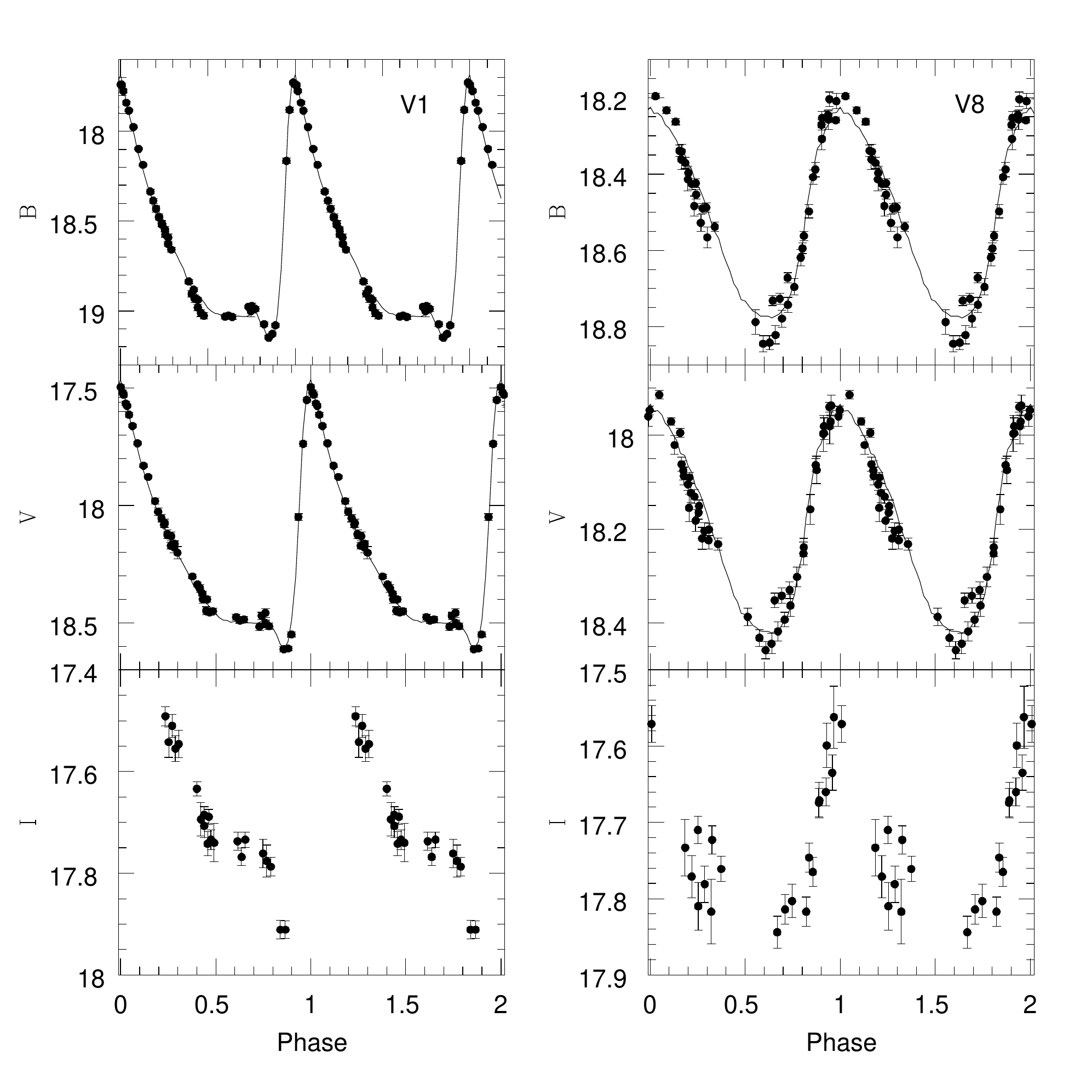}
     \caption{A sample of light curves showing an RRab light curve on the left and the only RRc on the right. All the light curves can be seen in the online edition of the paper.}
\end{figure}

\begin{figure}
     \includegraphics[width=\columnwidth]{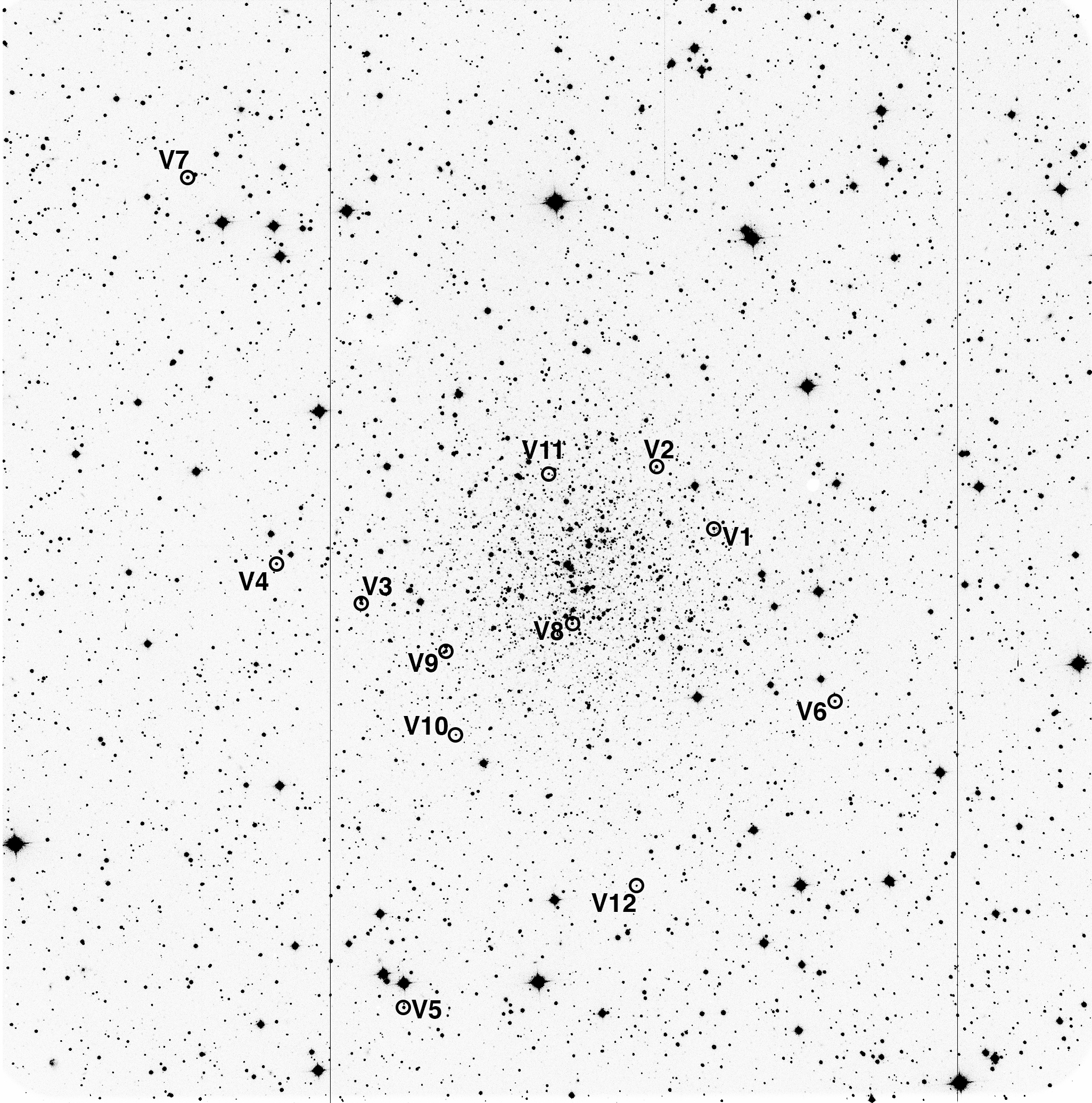}
     \caption{Finding chart for the RR~Lyrae stars in Arp~2. North is up and east is to the left. The field-of-view is approximately 13.7\arcmin x 13.7\arcmin. The online version of this figure can be used to zoom in on each individual variable.}
\end{figure}

\subsection{Gaia Data Release 2 and Catalina Surveys Southern Period Variable Star Catalogue}
Data for stars within the Arp~2 field were released in the Gaia Data Release 2\footnote{This work has made use of data from the European Space Agency (ESA) mission {\it Gaia} (\url{https://www.cosmos.esa.int/gaia}), processed by the {\it Gaia}
Data Processing and Analysis Consortium (DPAC, \url{https://www.cosmos.esa.int/web/gaia/dpac/consortium}). Funding for the DPAC has been provided by national institutions, in particular the institutions participating in the {\it Gaia} Multilateral Agreement.} (Gaia Collaboration et al.\ 2016, 2018). Of the variable stars in our field, V1, V3, V5, V6, V7, V8, and V9 were flagged as potential variables by Gaia. Of that subset, only V7, V8, and V9 had periods determined. These results are not completely unexpected since the number of observations in a given filter ranged from 9-18. We list the properties of the Arp~2 RRL as found in the Gaia database in Table~2. The magnitudes listed are through the $G$, $G_{BP}$ (Blue Photometer, wavelength range 330-680~nm), and $G_{RP}$ (Red Photometer,  wavelength range 640-1000~nm). The periods determined for the three RRL noted above match well with our observations. We also adjusted the magnitudes to the Johnson-Cousins system using the transformations listed in the appendix of Evans et al. (2018). Figure~4 shows the light curves for V7, V8, and V9 with the Gaia data as shifted to provide the best match with our light curves included. For the most part, the Gaia data fits in well with our photometry. As noted in Pancino et al.\ (2017), there have been some challenges when analyzing Gaia data in crowded fields. It is anticipated that the light curves from Gaia will improve as the mission continues to run.  

We also investigated the membership of the RRL stars detected in the Gaia database. Stars were selected from a region within 15\arcmin around Arp~2 to make sure we were sampling enough stars (the tidal radius was noted in Section~3). The other conditions we placed on the dataset were to have errors less than 0.15 milliarcseconds (mas) per year for both proper motions. This resulted in 760 stars being selected. In plotting the data according to the proper motions in right ascension and declination, we found a grouping of stars around $-2.5$ mas per year and $-1.6$ mas per year, respectively. Figure~5 plots the stars around this concentration. We added the RRL stars to the plot along with their error bars. The typical error bars for the stars in that region are shown offset to the right of the grouping. Given we limited our search using the uncertainties in right ascension and declination, it is not surprising to find that the typical error bars are smaller than those for the RRL stars. The figure shows that we cannot exclude any of those RRL stars found in the Gaia database from membership within Arp~2. We anticipate an improvement with this data in future Gaia data releases. That data will hopefully clear up membership questions on variable stars such as V7.

 \begin{table*}
  \caption{Arp~2 Variable Star Data from Gaia Database}             
  \label{table:2}      
  \centering          
  \begin{tabular}{c c c c c c}     
  \hline\hline       
  ID  & Gaia ID     & Period & $\overline{G}$ & $\overline{G_{BP}}$ & $\overline{G_{RP}}$ \\
        &                  & (days) \\ 
  \hline                    
     1 & 6746357023072172032 & \ldots                                        & 18.086 & 18.204 & 17.608 \\  
     2 & 6746379563060514432 & \ldots                                        & 17.959 & 18.220 & 17.482 \\
     5 & 6746352625025357696 & \ldots                                        & 17.929 & 18.185 & 17.475 \\
     6 & 6746356885633098240 & \ldots                                        & 18.112 & 18.229 & 17.730 \\
     7 & 6746382655437208192 & 0.5284960$\pm$0.0000014     & 17.628 & 17.927 & 17.096 \\
     8 & 6746380220188778880 & 0.29136053$\pm$0.00000061 & 18.118 & 18.191 & 17.665 \\
     9 & 6746379532994002048 & 0.5152797$\pm$0.0000010     & 17.987 & 17.914 & 17.400 \\
  \hline                  
  \end{tabular}
  \end{table*}

\begin{figure}
     \includegraphics[width=\columnwidth]{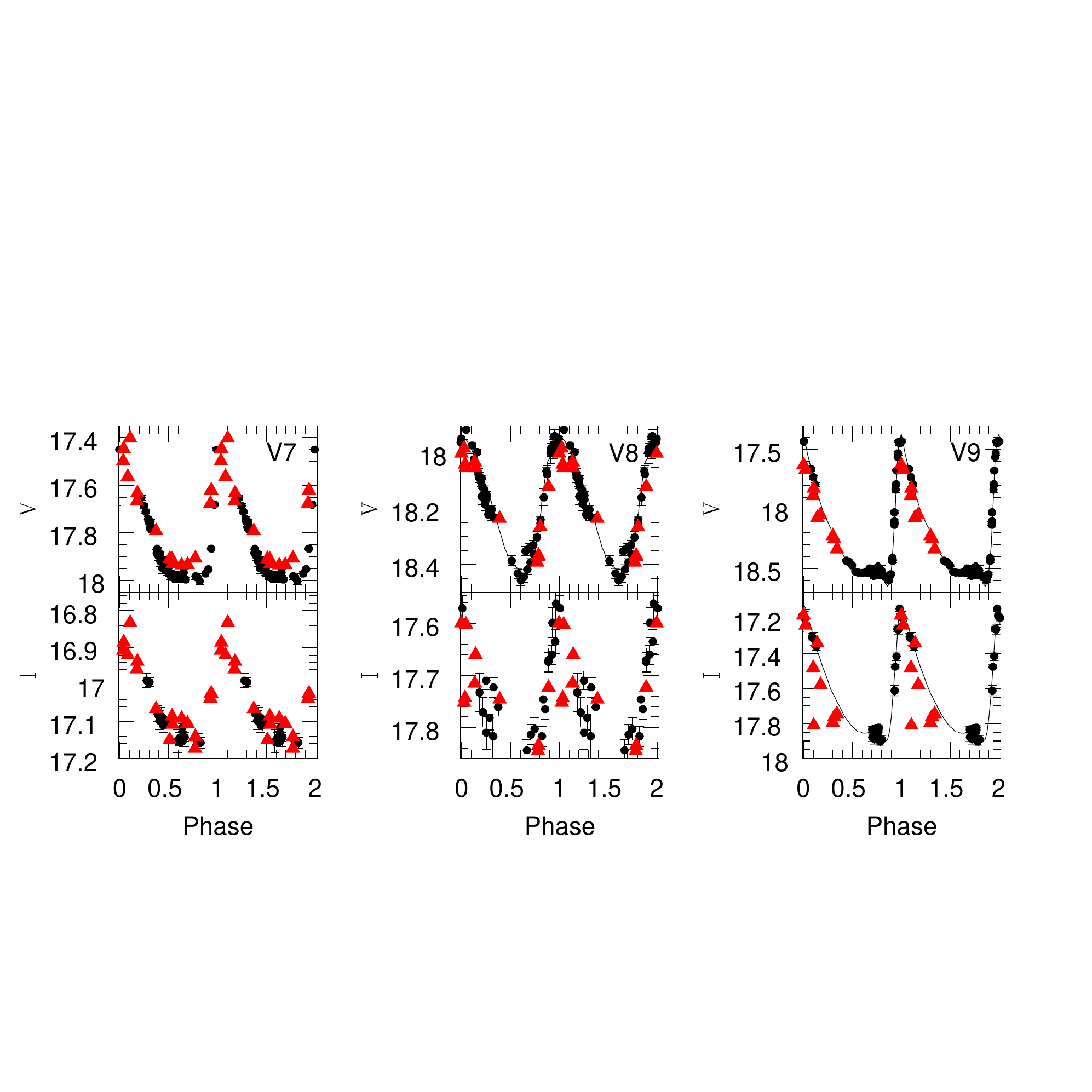}
     \caption{Light curves for V7, V8, and V9 (black circles; this paper) including the Gaia photometry (red triangles). The Gaia magnitudes were adjusted to the Johnson-Cousins system using the transformations listed in the appendix of Evans et al. (2018).}
\end{figure}

\begin{figure}
    \includegraphics[width=\columnwidth]{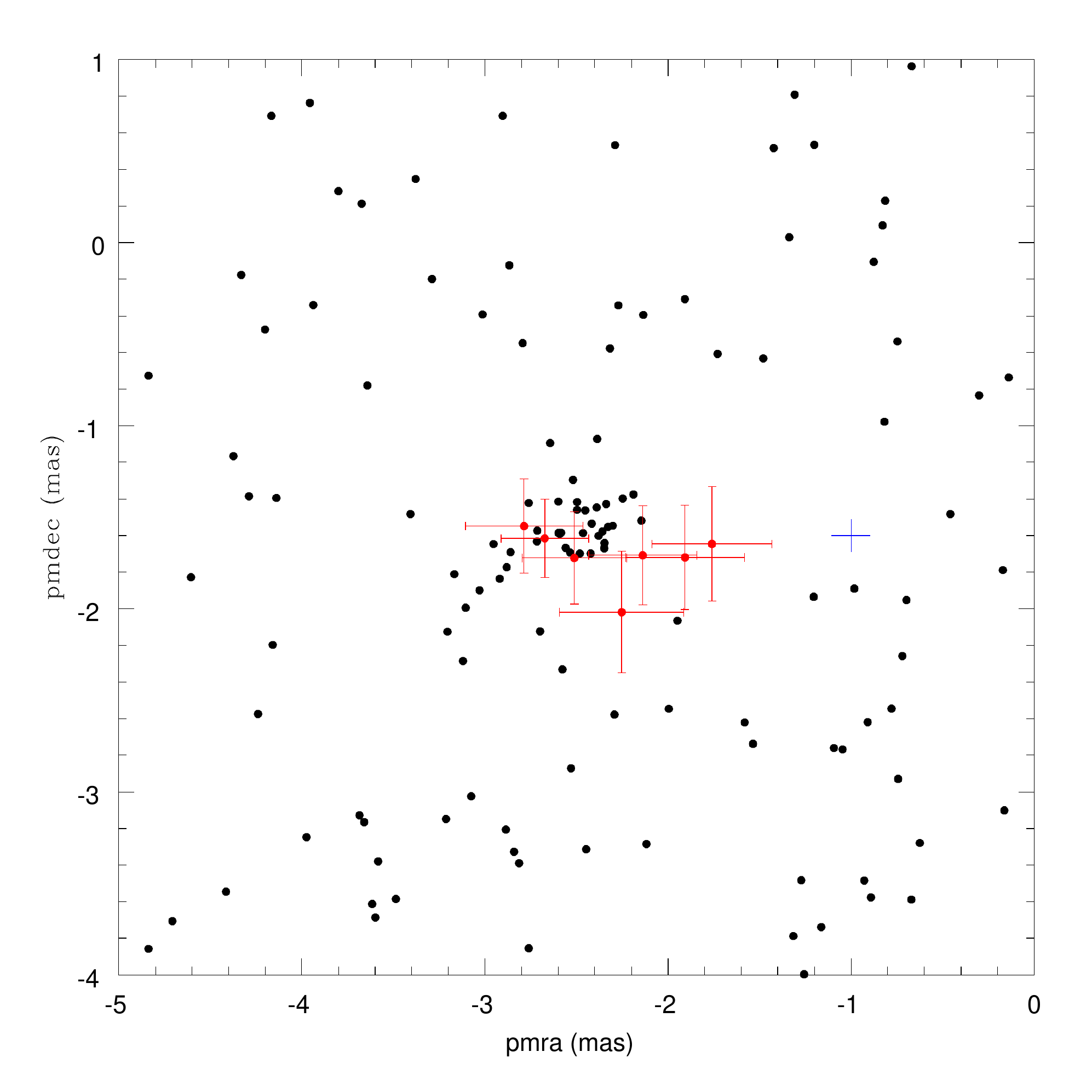}
    \caption{Comparison of the Gaia proper motions as measured in the right ascension and declination directions. RR~Lyrae stars are shown with error bars (in red). The smaller cross offset to the right of the RRL stars shows the typical error bars for stars within the grouping of stars at that region.}
\end{figure}

We cross-checked the Arp~2 variables stars with those found in the Catalina Surveys Southern periodic variable star catalogue, specifically Drake et al.\ (2017). Three RRL stars were found to match as shown in Table~3. The periods are similar to those found in our paper. However, the $V$-magnitudes are noticeably brighter than those found through the analysis of our and the OGLE-IV data. It should be noted that the Catalina survey does not use a standard Johnson $V$ filter and has no color term to aid in calibration. The Catalina survey $V$-amplitudes are also smaller than those found in Table~1. The smaller amplitudes, as well as the brighter Catalina magnitudes, may indicate that the Catalina photometry is influenced by blending in the crowded Arp~2 field.

 \begin{table*}
  \caption{Arp~2 Variable Star Data from Catalina Surveys Southern Periodic Variable Star Catalogue}             
  \label{table:3}      
  \centering          
  \begin{tabular}{c c c c c}     
  \hline\hline       
  ID  & Catalina ID     & Period & $\overline{V}$ & $A_{V}$ \\
        &                  & (days) \\ 
  \hline                    
     1 & SSS-J192836.2-302057 & 0.568617 & 17.794 & 0.991 \\  
     3 & SSS-J192856.7-302130 & 0.564851 & 16.947 & 0.561 \\
     9 & SSS-J192852.1-302212 & 0.515278 & 16.582 & 0.371 \\
  \hline                  
  \end{tabular}
  \end{table*}

\section{Analysis} \label{sec:analysis}

From the RRab stars in Arp~2 we find a mean period of $\langle P_{ab}\rangle = 0.581\pm0.047$ days, where the error is the standard error of the mean. Since there is only one RRc star, we do not list the mean period for that type of star. The RRab mean period combined with the metallicity places Arp~2 at the lower edge of the Oosterhoff gap (see Catelan \& Smith 2015). It borders between the Oosterhoff intermediate clusters and Oosterhoff~I clusters. We note that if V5 and V7 are left out of the mean period calculation because of the possibility that they may not be members (see Section~3), we find a RRab mean period of $\langle P_{ab}\rangle = 0.561\pm0.054$ days. This mean period would place Arp~2 among the Oosterhoff~I clusters, but the high standard deviation still makes its classification uncertain. The difficulty in classifying Arp~2 by Oosterhoff type largely arises from its relatively small number of RRab stars, particularly if the two stars of uncertain membership are excluded.

Some of the other globular clusters that are thought to be associated with the Sgr dwarf are as follows: NGC~6715 (M54; [Fe/H]$=-1.58$, $\langle P_{ab}\rangle = 0.599$~days, Hamanowicz et al.\ 2016), Terzan~8 ([Fe/H]$~-2.3$, $\langle P_{ab}\rangle = 0.644$, Salinas et al.\ 2005), NGC~4147 ([Fe/H]$=-1.83$, $\langle P_{ab}\rangle = 0.524$, Arellano Ferro et al.\ 2018), and NGC~5634 ([Fe/H]$=-1.88$, $\langle P_{ab}\rangle = 0.672$~days, Salinas et al.\ 2005). NGC~5634 and Terzan~8 are both clearly Oosterhoff~II clusters, while NGC~4147 is an Oosterhoff~I cluster. 

In Figure~6 we show the period-amplitude diagram for the RRL stars in Arp~2. Although Smith, Catelan, \& Kuehn (2011) questioned how accurately such a diagram can provide an Oosterhoff classification, it can still give us a general idea of trends when compared to other clusters. In each of the diagrams, there is scatter among the RRab stars. Most of them are near the Oosterhoff~I line, but there are a few scattered around the Oosterhoff~II line as well. 

\begin{figure}
   \centering
   \includegraphics{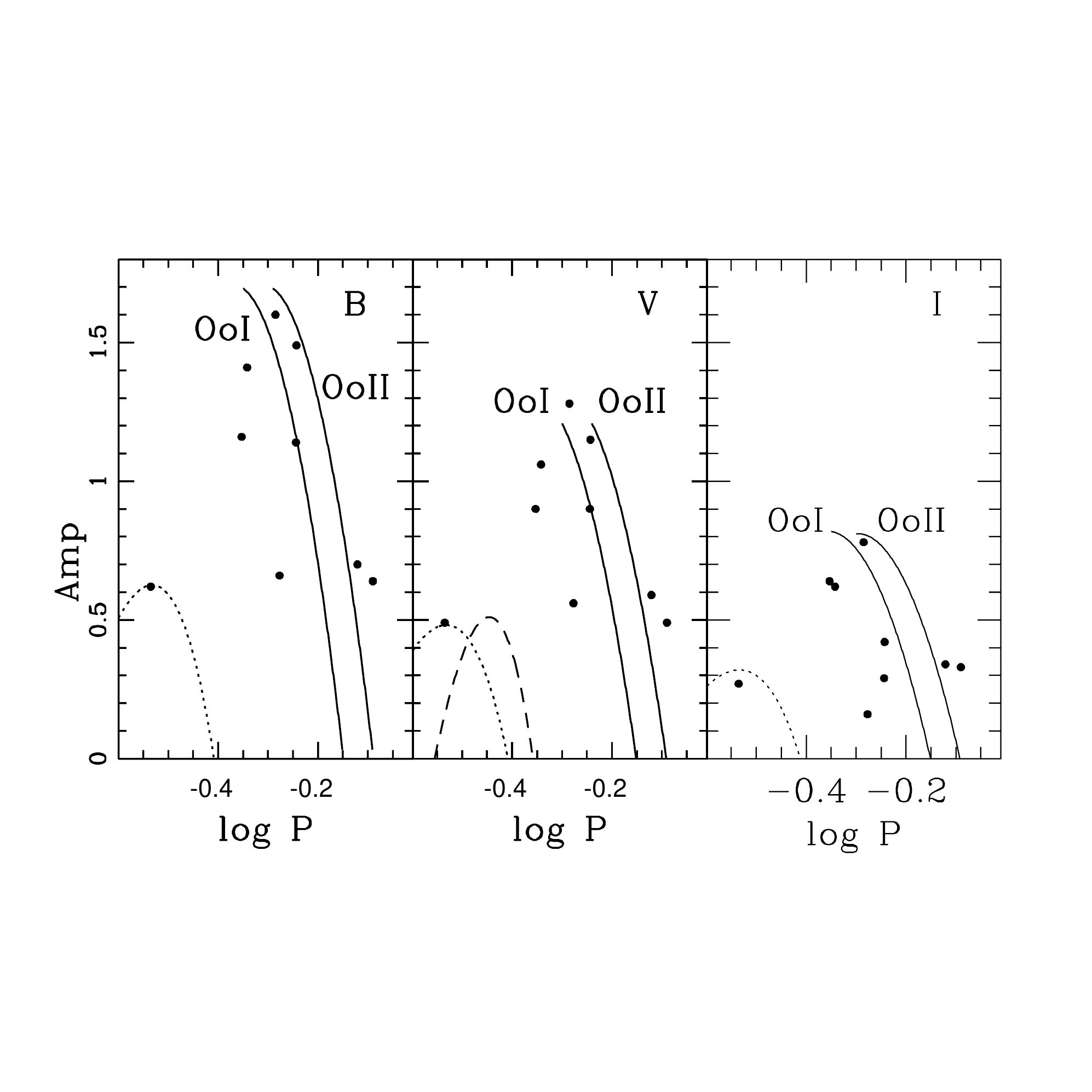}
   \caption{Period-amplitude diagram for RRL in Arp~2. In the $B$-data panel, the solid lines shown for the typical locations of the Oosterhoff~I and Oosterhoff~II RRab stars are taken from Zorotovic et al.\ (2010). The dotted line for the RRc stars was taken from the $V$-data and adjusted by a factor of 1.3. For the $V$-data we used the Zorotovic et al.\ lines. The dotted line represents the typical location for RRc stars in Oosterhoff~I clusters from Arellano Ferro et al.\ (2015) and the dashed line is for the Oosterhoff~II RRc stars taken from Kunder et al.\ (2013b). In the right panel, the lines shown for the RRab stars are from Kunder et al.\ (2013a). The dotted line for the OoI RRc stars comes from Arellano Ferro et al. (2015).}
\end{figure}
   
In using the RRL stars to estimate the distance to Arp~2, we calculated a mean magnitude of $V(RR) = 18.10\pm0.04$. We determined the absolute magnitude for the RRL stars using the relation from Catelan \& Cort\'{e}s (2008),
   \begin{equation}
      M_{V(RR)} = (0.23\pm0.04){\rm [Fe/H]}+(0.984\pm0.127).
   \end{equation}
With an adopted metallicity of [Fe/H]$=-1.83\pm0.07$ (Mottini, Wallerstein, \& McWilliam 2008) we find a value of $M_{V(RR)}=0.56\pm0.15$, where the error is the standard error of the mean. Harris (1996) lists the reddening toward Arp~2 as $E(B-V)=0.10$. Using $A_V=3.1E(B-V)$, we determined the distance modulus to be $(m-M)_0=17.24\pm0.17$. The uncertainty for the calculated value was determined from the standard propagation of errors.

We also used the relations from Catelan, Pritzl, \& Smith (2004) to determine the $I$-band absolute magnitudes for the RRL stars. The absolute magnitude for the RRL stars in the $I$-band are related to the period of each variable and the metallicity of the cluster. We used [Fe/H]$=-1.83\pm0.07$ and [$\alpha$/Fe]$=0.31\pm0.11$ to determine log$Z=-3.37$. Individual $M_I$ values were found for each RRL star. These values were subtracted from the individual $m_I$ values to determine an average of $(m_I-M_I)=17.48\pm0.07$ where the uncertainty is the standard error of the mean. The variable V7 had a noticeably brighter $I$-magnitude. If we leave that star out we find a mean value of $(m_I-M_I)=17.54\pm0.04$. Using $A_I=0.64A_V$ (Thoraval et al.\ 1997), we find $(m-M)_0=17.28\pm0.09$ including V7 and $(m-M)_0=17.34\pm0.07$ without V7, where the uncertainty is found through the standard propagation of the errors. The values are fainter than the value found using the $V$-data, but well within the uncertainty of the previous value.

An alternative distance modulus calculation is the Wesenheit method, which purports to be reddening free and metallicity independent. The method has primarily been applied to Cepheid variables (e.g., Turner 2010). For our RRL stars, we follow the relation from Kov\'{a}cs (2003). Given the poorer quality of the $I$-band light curves, we decided to use only the $V$ and $B$ data to derive the Wesenheit indices for the RRL stars. We plotted $W = V - 3.1 (B-V)$ versus log~$P$ for our RRL, after shifting the log of the period for V8 by 0.127 (e.g., Coppola et al.\ 2015) to fundamentalize it. We found a slope of $-2.45$, excluding V2 and V7 given they are clearly offset from the rest of the RRL. This matches well with the slope found by Kov\'{a}cs. To determine the distance modulus for the RRL stars in Arp~2, we rewrite the relationship related to the $B$ and $V$ magnitudes from Kov\'{a}cs to be

   \begin{equation}
      (m-M)_0 = V - 3.1 (B-V) + 2.467 log P_0 + 1.087
   \end{equation}

Table~4 gives the distance modulus for each RRL star based on the data in Table~1. Note that V2 and V7 clearly stand apart from the other RRab as being much brighter. As noted in Section~3, V2 is brighter than most of the RRL and has a notably longer period. We also discussed how V7 is outside the Arp~2 tidal radius and has unusually red colors. The Wesenheit index method thus further brings their membership into question. Excluding these two RRab stars, we find a mean distance modulus of $17.29\pm0.06$~mag, where the uncertainty is the standard error of the mean. This value matches well with the other distance moduli discussed above.

\begin{table*}
  \caption{Wesenheit Method Distance Moduli for Arp~2 RR~Lyrae}             
  \label{table:4}      
  \centering          
  \begin{tabular}{c c }     
  \hline\hline       
  ID  & $(m-M)_0$ \\
  \hline                    
     1 & 17.43 \\  
     2 & 16.64 \\
     3 & 17.01 \\
     4 & 17.49 \\
     5 & 17.31 \\
     6 & 17.24 \\
     7 & 16.43 \\
     8 & 17.21 \\
     9 & 17.33 \\
  \hline                  
  \end{tabular}
  \end{table*}

More recently, Braga et al.\ (2016) applied the Wesenheit method to the RRL stars of $\omega$~Centauri. In putting together both the RRab and RRc variables, they found a coefficient for the color term of 3.06 and a slope of $-2.574\pm0.112$.  If we use these values with the zero point from Kov\'{a}cs (2003) we find a mean value for the distance modulus of $17.27\pm0.05$~mag, where the uncertainty is the standard error of the mean. The value is in line with the value we found using the Kov\'{a}cs equation.

To determined the distance modulus from the SX~Phe stars, we have assumed all of stars are pulsating in the fundamental mode. Due to the poorer quality of these light curves, we are not able to clearly distinguish their pulsation modes. We used the period-luminosity relation for fundamental mode SX~Phe from McNamara (2011) to determine the distance modulus from these stars. The absolute magnitude through the $V$-filter was determined for each SX~Phe star, subtracted from the apparent visual magnitude, and corrected for reddening. The distance modulus was determined to be $(m-M)_0=17.27\pm0.04$, where the uncertainty is the standard error of the mean. The value is consistent with that found for the RRL stars in our study.

Buonanno et al.\ (1995) found $V_{HB}=18.30\pm0.07$. This is fainter than our {\bf $\langle V \rangle$} value for the RRL stars by about 0.16~mag, when excluding the two brighter RRL. Partly this is attributable to our photometry being 0.04~mag brighter than that of Buonanno et al. However, we also find this to be caused by a small slope in the brightness of the HB, noticeable when the RRL stars are included with the nonvariable HB stars in the CMD. The HB becomes slightly brighter as we move from the blue side of the instability strip (mainly used by Buonanno et al.) to the red side. Dotter et al.\ (2010) found the magnitude of the HB through the F606W filter to be 18.05~mag. This is much closer to our value, although it is through a different filter. Siegel et al.\ (2011) found $(m-M)_0=17.37$~mag. This is slightly larger than we found using the visual magnitudes of the RRL, but is similar to what we found using the $I$-band method without V7.

\subsection{Reddening}

We have used the light curves of the RRab stars to investigate the reddening toward Arp~2. From the minimum light in the phase of 0.5-0.8, we used the methods described in Blanco (1992), which originated from Sturch (1966), for $(B-V)_{min}$ magnitudes and Guldenschuh et al. (2005) for the $(V-I)_{min}$ magnitudes. Specifically, the equations we used are listed below.

   \begin{equation}
      E(B-V) = (B-V)_{min} - 0.24P - 0.056[Fe/H]_{ZW84} - 0.336
   \end{equation}

   \begin{equation}
      E(V-I) = (V-I)_{min} - 0.58\pm0.02
   \end{equation}
   
In Table~5 we list the minimum colors of the RRab stars along with their corresponding color excesses. We adopted the same metallicity as used in the previous section. From the ($B-V$) colors, we find an average color excess of E($B-V$)$=0.16\pm0.02$, where the uncertainty is the standard error of the mean. The value is notably higher than that from Harris (1996). From the ($V-I$) colors, we find an average color excess of E($V-I$)$=0.13\pm0.04$. Recent values of E(V-I)/E(B-V) in the literature show some spread, and the actual value will also depend with the adopted value of $R_V$ (e.g. Inno et al. 2016, Schlgel et al. 1998, Schlafly \& Finkbeiner 2011, Guldenschuh et al. 2005). We have adopted the value of 1.35 from Kunder et al. (2013b), which is an average value from several sources noted within the paper. Using the color excess ratio of E($V-I$)/E($B-V$)$=1.35$, we find a color excess of E($B-V$)$=0.10\pm0.04$, where the uncertainty follows the standard propagation of errors. This value matches the Harris value well. It should be noted that the RRab stars, V2 and V5, have light curves that are still declining through the 0.5-0.8 phase rather than being flat at minimum light.

The reddening calculators in the NASA/IPAC Infrared Science Archive\footnote{http://irsa.ipac.caltech.edu/applications/DUST/} allow us to calculate the reddening in the direction toward each RR Lyrae star, using the methods of Schlafly \& Finkbeiner (2011) and Schlegel, Finkbeiner, \& Davis (1998). For the nine RR Lyrae stars, we get a mean value of E($B-V$) $= 0.097\pm0.005$ from the  Schlafly \& Finkbeiner method and $0.113\pm0.005$ from the Schlegel, Finkbeiner, \& Davis method, where the uncertainties are the standard error of the mean. Those results are in good agreement with the reddening of E($B-V$) = 0.11 derived by Buonanno et al.\ (1995) based on the method outlined in Burstein \& Heiles (1982).  Compared to the reddening values in Table 5, the most discrepant result is for V7, for which we derived E($B-V$) = 0.28 and E(V-I) = 0.27 (equivalent to E($B-V$) = 0.20).  In the direction of V7, the Schlafly \& Finkbeiner value is E($B-V$) = 0.105 and the Schlegel, Finkbeiner, \& Davis value is E($B-V$) = 0.123.  We have already noted the unusually red colors for V7 in our photometry.  If V7 is excluded from our averages of the reddening values in Table 5, the average values become E($B-V$) $= 0.14\pm0.01$ from the $B-V$ observations and E($B-V$) $= 0.08\pm0.02$ from the $V-I$ observations of the remaining seven RRab variables.

 \begin{table*}
  \caption{Arp~2 RRab Color Excess Determinations}             
  \label{table:5}      
  \centering          
  \begin{tabular}{c c c c c }     
  \hline\hline       
  ID  & $(B-V)_{min}$ & E($B-V$) & $(V-I)_{min}$ & E($V-I$) \\
  \hline                    
     1 & 0.52                & 0.15     & 0.73                & 0.15   \\  
     2 & 0.57                & 0.14     & 0.77                & 0.19   \\
     3 & 0.54                & 0.17     & 0.75                & 0.17   \\
     4 & 0.46                & 0.12     & 0.63                & 0.05   \\
     5 & 0.54                & 0.12     & 0.67                & 0.09   \\
     6 & 0.49                & 0.15     & 0.60                & 0.02   \\
     7 & 0.64                & 0.28     & 0.85                & 0.27   \\
     9 & 0.50                & 0.14     & 0.67                & 0.09   \\
  \hline                  
  \end{tabular}
  \end{table*}

\section{Summary} \label{sec:sum}

We presented the properties of the variable stars found in Arp~2. It is a key globular cluster to study given its former association with the Sagittarius dwarf galaxy. Arp~2 is also of interest given its younger age while being relatively metal-poor. Of the nine detected RRL stars, eight are of RRab type and one is RRc type. The mean period of the RRab stars is $\langle P_{ab}\rangle = 0.581\pm0.047$~days, which places Arp~2 at the edge of the Oosterhoff gap on the Oosterhoff~I side. Excluding two possible non-members, the mean RRab period becomes $0.561\pm0.054$~days. We examined the Gaia database to see what was found for the Arp~2 variable stars. Of the nine variable stars, seven were flagged as variable stars by Gaia. Of those, only three had periods determined. Those periods matched well with the ones we determined. We investigated the Gaia proper motions and could not definitively exclude any of the RRL stars as being members of Arp~2. From the RRL stars in our study we found a distance modulus of $(m-M)_0=17.24\pm0.17$ using the $V$-band and $(m-M)_0=17.28\pm0.09$ using the $I$-band, when all of the RRL are included. Using Wesenheit indices, we determined a distance modulus of $(m-M)_0=17.29\pm0.06$. From the SX~Phe stars, we found a distance modulus of $(m-M)_0=17.27\pm0.04$. The properties we determined from the variable stars do not cause Arp~2 to stand out in any significant way from the Milky Way GCs, which is likely why it was not seen as being an extragalactic GC until the discovery of the Sgr dwarf galaxy.

\acknowledgments

Support for B.J.P.\ was given by the University of Wisconsin Oshkosh Faculty Development Grant (FDR 953, FDR 1059). Support for M.C.\ is provided by the Ministry for the Economy, Devolpment, and Tourism's Millennium Science Initiative through grant IC\,120009, awarded to the Millennium Institute of Astrophysics (MAS); by Proyecto Basal PFB-06/2007; by CONICYT's PCI program through grant DPI20140066; and by FONDECYT grant \#1171273. Thank you to A. Layden for his assistance with his light curve analysis program.





\begin{thebibliography}{}

\bibitem[Anderson et al.\ (2008)]{anderson08} Anderson, J., et al.\ 2008, \aj, 135, 2055
\bibitem[Arellano Ferro et al.\ (2015)]{arellano15} Arellano Ferro, A., Mancera, Pi\~{n}a, P. E., Bramich, D.M., Giridhar, S., Ahumada, J. A., Kains, N., \& Kuppuswamy, K. 2015, \mnras, 452, 727
\bibitem[Arellano Ferro et al.\ (2018)]{arellano18} Arellano, Ferro, A., Rojas Galindo, F. C., Muneer, S., \& Giridhar, S. 2018, Revista Mexicana de Astronom\'{i}a y Astr\'{o}f\'{i}sica, (arXiv: 1803.05313)
\bibitem[Blanco(1992)]{blanco92}  Blanco, V.~M.\ 1992, \aj, 104, 734
\bibitem[Braga et al.\ (2016)]{braga16} Braga, V.~F., et al.\ 2016, \aj, 152, 170
\bibitem[Buonanno et al.\ (1995)]{buonanno95} Buonanno, R., Corsi, C. E., Pecci, F. F., Richer, H. B., \& Fahlman, G. G. 1995, \aj, 109, 650
\bibitem[Burstein \& Heiles (1982)]{burstein82} Burstein, D., \& Heiles, C. 1982, \aj, 87, 1165
\bibitem[Carraro \& Seleznev (2011)]{carraro11} Carraro, G., \& Seleznev, A. F. 2011, \mnras, 412, 1361
\bibitem[Carraro et al.\ (2007)]{carraro07} Carraro, G., Zinn, R., \& Moni Bidin, C. 2007, \aap, 466, 181
\bibitem[Catelan \& Cort\'{e}s (2008)]{catelan08} Catelan, M., \& Cort\'{e}s, C. 2008, \apj, 676, 135
\bibitem[Catelan et al.\ (2004)]{catelan04} Catelan, M., Pritzl, B. J., \& Smith, H. A. 2004, ApJSS, 154, 633
\bibitem[Catelan \& Smith (2015)]{catelan15} Catelan, M., \& Smith, H. A. 2015, Pulsating Stars, Wiley-VCH
\bibitem[Coppola et al.\ (2015)]{coppola15} Coppola, G., et al.\ 2015, \apj, 814, 71
\bibitem[Dotter et al.\ (2010)]{dotter10} Dotter, A., et al.\ 2010, \apj, 708, 698
\bibitem[Evans et al.\ (2018)]{evans18} Evans, D. W., et al.\ 2018, \aap, (arXiv:1804.09368)
\bibitem[Font et al.\ (2011)]{font11} Font, A. S., McCarthy, I. G., Crain, R. A., Theuns, T., Schaye, J., Wiersma, R. P. C., Dalla Vecchia, C. 2011, \mnras, 416, 2802
\bibitem[Gaia Collaboration et al.\ (2016)]{gaia16} Gaia Collaboration, et al.\ 2016, \aap, 595, A1
\bibitem[Gaia Collaboration et al.\ (2018)]{gaia18} Gaia Collaboration, et al.\ 2018, \aap, (arXiv:1804.09379)
\bibitem[Guldenschuh et al.(2005)]{2005PASP..117..721G}  Guldenschuh, K.~A., et al.\ 2005, \pasp, 117, 721
\bibitem[Harris (1996)]{harris96} Harris, W. E. 1996, \aj 112, 1487
\bibitem[Hamanowicz et al.\ (2016)]{hamanowicz16} Hamanowicz, A., et al.\ 2016, Acta Astronomica, 66, 197
\bibitem[Ibata et al.\ (1994)]{ibata94} Ibata, R. A., Gilmore, G., \& Irwin, M. J. 1994, Nature, 370, 194
\bibitem[Ibata et al.\ (1995)]{ibata95} Ibata, R. A., Gilmore, G., \& Irwin, M. J. 1995, \mnras, 277, 781
\bibitem[Inno et al.\ (2016)]{inno16} Inno, L., Bono, G., \& Matsunaga, N. 2016, \apj, 832, 176
\bibitem[Kov\'{a}cs (2003)]{kovacs03} Kov\'{a}cs, G. 2003, \mnras, 342, L58
\bibitem[Kov\'(a)cs \& Kupi (2007)]{kovacs07} Kov\'{a}cs, G., \& Kupi, G. 2007, \aap, 462, 1007
\bibitem[Kunder et al.\ (2013a)]{kunder13a} Kunder, A., Stetson, P. B., Catelan, M., Walker, A. R., \& Amigo, P. 2013a,  \aj, 145, 33
\bibitem[Kunder et al.\ (2013b)]{kunder13b} Kunder, A., Stetson, P. B., Cassisi, S., et al.\ 2013b, \aj, 146, 119
\bibitem[Layden \& Sarajedini (2000)]{layden00} Layden, A. C., \& Sarajedini, A. 2000, \aj, 119, 1760
\bibitem[McNamara (2011)]{mcnamara11} McNamara, D. H. 2011, \aj, 142, 110
\bibitem[Mottini et al.\ (2008)]{mottini08} Mottini, M., Wallerstein, G., \& McWilliam, A. 2008, \aj, 136, 614
\bibitem[Oosterhoff (1939)]{oosterhoff} Oosterhoff, P. T. 1939, Observatory, 62, 104
\bibitem[Pancino et al.\ (2017)]{pancino17} Pancino, E., Bellazzini, M., Giuffrida, G., \& Marinoni, S. 2017, \mnras, 467, 412
\bibitem[Salinas et al.\ (2005)]{salinas05} Salinas, R., Catelan, M., Smith, H. A., Pritzl, B. J., \& Borissova, J. 2005, IBVS, 5640, 1
\bibitem[Salinas et al.\ (2012)]{salinas12} Salinas, R., J\'{i}lkov\'{a}, L., Carraro, G., Catelan, M., \& Amigo, P. 2012, \mnras, 421, 960
\bibitem[Schlafly \& Finkbeiner (2011)]{schlafly11} Schlafly, E.~F., \& Finkbeiner, D.~P. 2011, \apj, 737, 103
\bibitem[Schlegel, Finkbeiner, \& Davis (1998)]{schlegel98} Schlegel, D.~J., Finkbeiner, D.~P., \& Davis, M. 1998, \apj, 500, 525
\bibitem[Siegel et al.\ (2011)]{siegel11} Siegel, M. H., et al.\ 2011, \apj, 743, 20
\bibitem[Smith, Catelan, \& Kuehn (2011)]{smith} Smith, H. A., Catelan, M., Kuehn, C. 2011, in Proc. Carnegie Astrophys. Ser., Vol. 5, RR Lyrae Stars, Metal-Poor Stars, and the Galaxy (Washington, DC: Carnegie Institute of Washington), 17
\bibitem[Sohn et al.\ (2018)]{sohn18} Sohn, S.~T., Watkins, L.~L., Fardal, M.~A., van der Marel, R.~P., Deason, A.~J., Besla, G., 
\& Bellini, A.\ 2018, \apj, 862, 52
\bibitem[Stellingwerf (1978)]{stellingwerf} Stellingwerf, R. F. 1978, \pasp, 108, 851
\bibitem[Stetson (1987)]{stetson87} Stetson, P. B. 1987, \pasp, 99, 191
\bibitem[Stetson (1994)]{stetson94} Stetson, P. B. 1994, \pasp, 106, 250
\bibitem[Stetson (2000)]{stetson00} Stetson, P. B. 2000, \pasp, 112, 925
\bibitem[Sturch (1966)]{sturch66} Sturch, C. 1966, \apj, 143, 774
\bibitem[Thoraval, Boiss\'e, \& Duvert (1997)]{thoraval} Thoraval, S., Boiss\'e, P., \& Duvert, G. 1997, \aap, 319, 948
\bibitem[Turner (2010)]{turner10} Turner, D.~G.\ 2010, \apss, 326, 219
\bibitem[Valenti (2001)]{valenti01} Valenti, E., 2001, Ricerca delle variabili RR Lyrae neglie amassi globulari NGC 6304 e Arp 2, Tesi di Laurea (Universit\'{a} degli studi di Bologna, Bologna)
\bibitem[Zorotovic et al.\ (2010)]{zorotovic} Zorotovic, M., Catelan, M., Smith, H. A., et al.\ 2010, \aj, 139, 357

\end{thebibliography}
\end{document}